\documentclass[preprint,12pt]{elsarticle}

\usepackage{amssymb}
\usepackage{amsmath}
\biboptions{sort&compress}
\usepackage{bm}

\begin{document}

\begin{frontmatter}



\title{Experimental demonstration of picometer level signal extraction with time-delay interferometry technique}



\author[1]{Mingyang Xu}
\author[1]{ Yujie Tan\corref{mycorrespondingauthor}}
\cortext[mycorrespondingauthor]{Corresponding author.}
\ead{yjtan@hust.edu.cn}
\author[1]{Yurong Liang\corref{mycorrespondingauthor}}
\ead{liangyurong20@hust.edu.cn}
\author[1]{Jiawen Zhi}
\author[1]{Xiaoyang Guo}
\author[1]{Dan Luo}
\author[1]{Panpan Wang\corref{mycorrespondingauthor}}
\ead{ppwang@hust.edu.cn}
\author[1,2]{Hanzhong Wu\corref{mycorrespondingauthor}}
\ead{wuhanzhong@hust.edu.cn}
\author[1]{Chenggang Shao\corref{mycorrespondingauthor}}
\ead{cgshao@hust.edu.cn}

\address[1]{MOE Key Laboratory of Fundamental Physical Quantities Measurements, Hubei Key Laboratory of Gravitation and Quantum Physics, PGMF and School of Physics, Huazhong University of Science and Technology, Wuhan 430074, China}
\address[2]{State Key Laboratory of Applied Optics, Changchun Institute of Optics, Fine Mechanics and Physics, Chinese Academy of Sciences, Changchun 130033, China}



\begin{abstract}
In this work, we have built an experimental setup to simulate the clock noise transmission with two spacecrafts and two optical links, and further demonstrated the extraction of picometer level signal drowned by the large laser frequency noise and clock noise with the data post-processing method. Laser frequency noise is almost eliminated by using the idea of time-delay interferometry (TDI) to construct an equal arm interferometer. Clock asynchronism and clock jitter noise are significantly suppressed by laser sideband transmitting the clock noise using an electro-optic modulator (EOM). Experimental results show a reduction in laser frequency noise by approximately $10^5$ and clock noise by $10^2$, recovering a weak displacement signal with an average amplitude about 60 picometer and period 1 second. This work has achieved the principle verification of the noise reduction function of TDI technique to some extent, serving the data processing research of space-borne gravitational wave detection.
\end{abstract}




\end{frontmatter}


\section{Introduction}
With the successful direct detection of gravitational wave (GW) by the Laser Gravitational-Wave Observatory (LIGO) in 2016 \cite{gw1}, people can study the universe in a unique and entirely new way. Ground-based GW detection is mainly sensitive to GWs in the frequency band of 10-$10^4$ Hz, which is limited by ground vibration noise and gravity gradient noise below 10 Hz. In order to find more abundant GW sources, space-borne GW detection mission has been proposed, including LISA \cite{lisa}, DECIGO \cite{decigo}, Tianqin \cite{tianqin}, Taiji \cite{taiji}, etc.

The space-borne GW detector, such as LISA, consists of three spacecrafts. Two drag-free test masses are housed in each spacecraft serving as inertial references. When GWs pass by, the distance between the test masses in two adjacent spacecraft changes slightly. Heterodyne laser interferometers are used to measure this tiny distance change, in which a clock is required to trigger the digital sampling process. The GW signals are so weak that a lot of noise will overwhelm them. Typically, the laser frequency noise and the clock noise are the two main noise sources, which are about 7 and 3 orders of magnitude higher than the typical GW signals, respectively. To suppress these noises from hardware is cost and difficult. Thus, a post-processing technology named time-delay interferometry (TDI) is proposed \cite{tdi_1999,tdi1,tdi_wpp_2021}. TDI uses the different combination of data streams to construct a virtual equal arm interferometry which can reduce laser frequency noise. To suppress the clock noise, an EOM is used to transfer local clock noise to remote spacecraft which can construct a clock comparison chain \cite{tdi_2001,tdi_2018,tdi_2021,tdi_wpp_2,yzj}. Another clock noise reduction strategy uses optical frequency comb to link laser and clock, and then one can simultaneously remove the laser and clock noises by modifying the TDI combination \cite{tdi_ofc_2015,tdi_ofc_2022}. 

Theoretical researches have shown the importance and necessity of TDI post-processing algorithm in space-borne GW detection. To demonstrate the aspects of the signal processing chain, several laboratory experiments are designed. The first demonstration experiment \cite{tdi_ex_prl} using the Sagnac interferometer showed that the laser frequency noise can be eliminated by using the laser round-trip data streams, and the clock noise can be eliminated by using the EOM sideband modulation. Other experiments have demonstrated the Michelson combination to suppress laser frequency noise using long fiber delay \cite{tdi_ex_ol,tdi_ex_taiji} or electronic-phase-delay unit delay \cite{tdi_ex_prd}, and recently, researchers use a hexagonal optical bench to verify clock synchronization scheme down to LISA performance levels between three satellites \cite{clocksy_2022}. Optical frequency comb based TDI experiments \cite{tdi_ofc_2020,tdi_ofc_ole,tdi_ofc_ol} also demonstrate post-processing algorithms to suppress laser frequency noise and clock noise. Researchers have made a lot of effort to suppress the interferometer noises and electronic noises, so these experimental results demonstrated the excellent performance of the TDI suppression technique. The previous works focus on the noise floor of the system \cite{tdi_ex_prl,tdi_ex_taiji,tdi_ex_prd,clocksy_2022} or the concept presentation \cite{tdi_ex_ol,tdi_ofc_2020,tdi_ofc_ole,tdi_ofc_ol}, however, they did not introduce a real displacement signal, and did not check whether the real picometer (pm) level signal could be extracted after the laser frequency noise and clock noise were suppressed.

In this work, we present a TDI like system for extracting pm level displacement signal under large laser frequency noise and clock noise. The results show that the post-processing technology reduces laser frequency noise and clock noise by $10^5$ and $10^2$, recovering a real displacement signal with average amplitude 60 pm driven by a piezoelectric (PZT) ceramic. This confirms that weak signals can be restored by post-processing even if they are drowned by large laser frequency noise and clock noise.

\section{Experiment setup}
\begin{figure}[ht]
\centering
{\includegraphics[width=12cm]{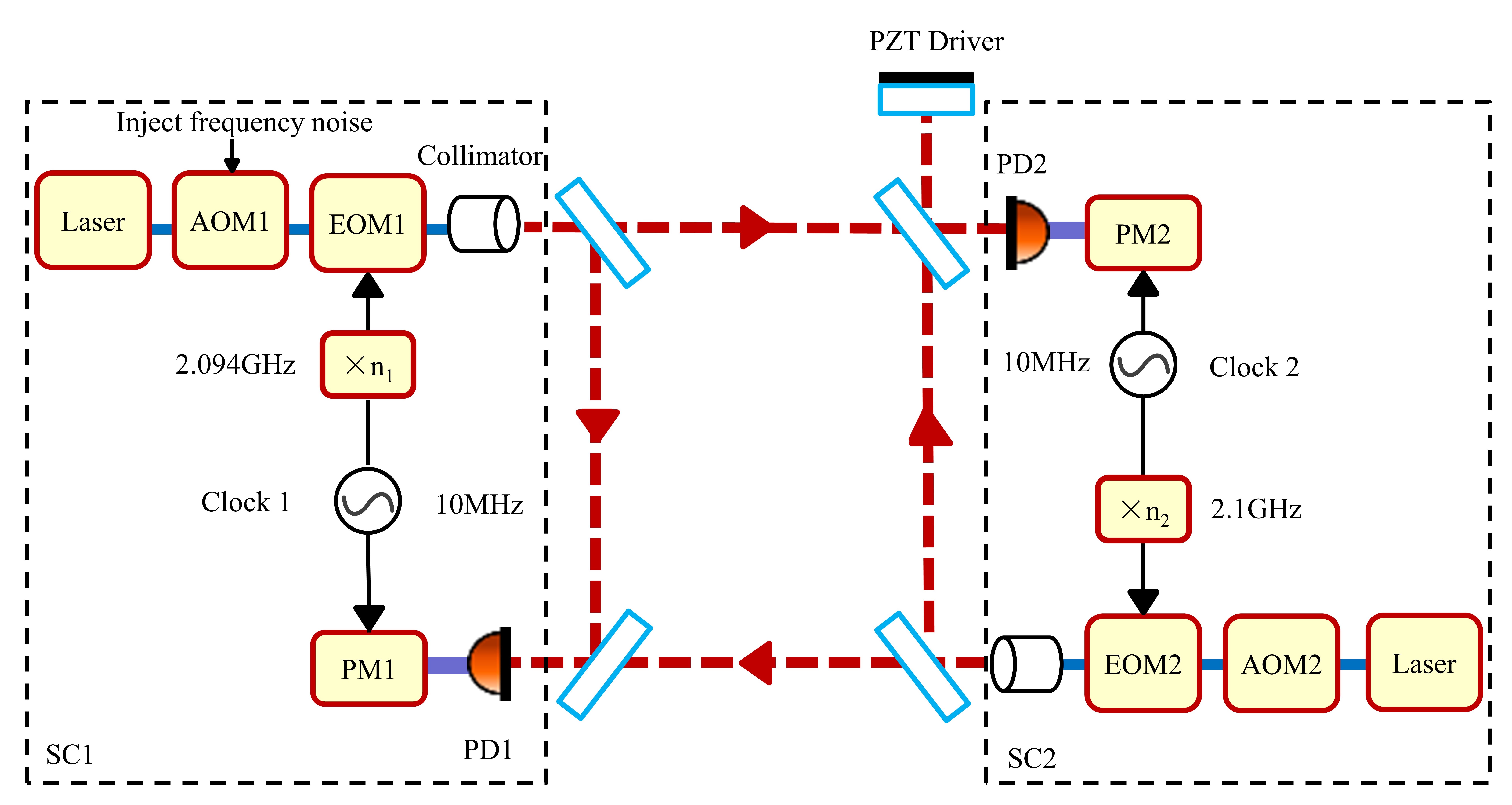}}
\caption{Schematic of the experimental setup to demonstrate the picometer level signal extraction under large laser frequency noise and clock noise. SC: spacecraft; PM: phasemeter; PD: photodetector.}
\label{fig:1}
\end{figure}

Fig. \ref{fig:1} shows the experimental setup. This experiment is set up to resemble laser interference measurements between two spacecrafts, and each spacecraft carries one laser, one independent clock, one phasemeter, and one clock transfer chain based on the EOM. A 1064 nm fiber laser (RIO) is split into two paths, one path pass through a fiber acousto-optic modulator (AOM, AAoptics-80 MHz) with 75 MHz driving frequency, the other pass through another AOM2 with 85 MHz driving frequency. In AOM1, a frequency noise is added to the driving frequency \cite{tdi_ofc_2020}. So, the laser frequency noises from AOM1 and AOM2 are independent and different. A square loop consisting of four beam splitters is designed to resemble the laser interference between two satellites. And a PZT-driven mirror is used to introduce a real weak displacement signal. The two laser interferences then can be received in the photodetectors (KEYANG), and the phase or frequency information can be obtained from homemade FPGA phasemeters \cite{liang}. Each phasemeter is triggered by an independent clock, so clock noise is introduced during the heterodyne measurements. To eliminate clock noise, a high frequency source (Rigol DSG821) is used to convert the 10 MHz clock to GHz, then the up-conversion clock noise is modulated to the sideband of the laser by the EOM (iXblue NIR-10G). In order to suppress the noise of the interferometer, the optical fiber section and the free optical path are placed separately in two thermal insulation boxes to minimize environmental disturbances (airflow and temperature fluctuations).

 The carrier data stream 1 and low sideband data stream 1 measured by the phasemeter 1 can be written as:
\begin{equation}
s_{1}^{c}\left( t_1 \right) =p_2\left( t_1 \right) -p_1\left( t_1 \right) -a_1{q}_1\left( t_1 \right), 
\label{eq1}
\end{equation}

\begin{align}
s_{1}^{sb}\left( t_1 \right) =&p_2\left( t_1 \right) -p_1\left( t_1 \right) -m_2{q}_2\left( t_1 \right) +m_1{q}_1\left( t_1 \right)\\ \nonumber
&-\left( a_1-m_2+m_1 \right) {q}_1\left( t_1 \right). 
\label{eq2}
\end{align}
Here, $p(t)$ is the laser frequency noise, ${q}(t)$ is the dimensionless relative clock noise, $a$ is the heterodyne interference frequency with $a_1=v_2-v_1=85$ $\rm{MHz}-75$ $\rm{MHz}=10$ $\rm{MHz}$, where $v_1$ is the center frequency of the laser after passing through AOM1 and $v_2$ is the center frequency of the laser after passing through AOM2. $m$ is the EOM modulation frequency.
Through the combination of carrier and sideband, one can extract expressions that mainly contain clock noise:
\begin{equation}
r_1(t_1)=s_{1}^{c}\left( t_1 \right) -s_{1}^{sb}\left( t_1 \right) =m_2{q}_2\left( t_1 \right) -m_2{q}_1\left( t_1 \right). 
\label{eq3}
\end{equation}
Similarly, the carrier data stream 2 and low sideband data stream 2 measured by the phasemeter 2 can be written as:
\begin{equation}
s_{2}^{c}\left( t_2 \right) =p_1\left( t_2 \right) -p_2\left( t_2 \right) -a_2{q}_2\left( t_2 \right)+h(t_2), 
\label{eq4}
\end{equation}
\begin{align}
s_{2}^{sb}\left( t_2 \right) =&p_1\left( t_2 \right) -p_2\left( t_2 \right) -m_1{q}_1\left( t_2 \right) +m_2{q}_2\left( t_2 \right)\\ \nonumber
&-\left( a_2-m_1+m_2 \right) {q}_2\left( t_2 \right)+h(t_2), 
\label{eq5}
\end{align}
where $a_2=-a_1$ and $h(t)$ is the displacement signal driven by the PZT, and the combination of carrier 2 and sideband 2 is:
\begin{equation}
r_2(t_2)=s_{2}^{c}\left( t_2 \right) -s_{2}^{sb}\left( t_2 \right) =m_1{q}_1\left( t_2 \right) -m_1{q}_2\left( t_2 \right). 
\label{eq6}
\end{equation}
It can be seen that the main noise of $r_1$ and $r_2$ is consistent, but the asynchronous of the clocks leads to residual noise in the subtraction of the two noises. The asynchronous of the clocks can be written as \cite{clocksy_2022}:
\begin{equation}
t_1\left( \tau \right) =t_2\left( \tau \right) +\delta \tau _{2,0}+\delta \tau _{2}. 
\label{eq7}
\end{equation}
where $\delta \tau _{2,0}$ is the constant initial time offset between clock1 and clock2, $\delta \tau _{2}$ is the timing noise of clock2 relative to the clock1.
Therefore, we need to synchronize the sampling time of the two clocks.  Taking clock1 as the primary clock, then, the difference of timing noise between clock2 and clock1 can be obtained from the data $r_2$ collected by clock2, which is:
\begin{equation}
\delta \tau _{2}\approx\int_{0}^{\tau} \frac{r_2}{m_1} d\tau. 
\label{eq111}
\end{equation}
And the constant initial time offset between clock1 and clock2 can be obtained by pseudo-random code ranging (PRNR) \cite{cqg_2011,yh} or time-delay Interferometry ranging (TDIR) \cite{tdiranging_2005}. In our experiment, we use the TDIR process which is also a post-processing technology. That is:
\begin{equation}
\gamma(\Lambda)=\frac{a_1}{m_2}r_1(t_1)+\frac{a_1}{m_1}r_2(t_2+\Lambda+\delta \tau _{2}). 
\label{eq11}
\end{equation}
By using dynamic fractional delayed interpolation \cite{clocksy_2022,tdi_cz,oe_2012}, we can translate $r_2$ on any time scale without introducing additional data flows. Then, the constant initial time offset $\delta \tau _{2,0}$ is determined by finding the minimum value of $\gamma(\Lambda)$ for different translation times $\Lambda$.
After clock synchronization, combining two carrier data streams can eliminate laser frequency noise:
\begin{equation}
s_{1}^{c}\left( t \right)+s_{2}^{c}\left( t \right) =h(t)-a_1(q_1\left( t \right)-q_2\left( t \right)). 
\label{eq8}
\end{equation}
At this time, the clock noise dominates, and then through the combination of the carrier and sideband data streams, we can further eliminate the clock noise while the signal remains:
\begin{equation}
\eta=s_{1}^{c}\left( t \right) +s_{2}^{c}\left( t \right)-\frac{a_1}{m_2}r_1(t)=h(t).
\label{eq9}
\end{equation}

\section{Experiment results}
 \begin{figure}[ht]
\centering
{\includegraphics[width=12cm]{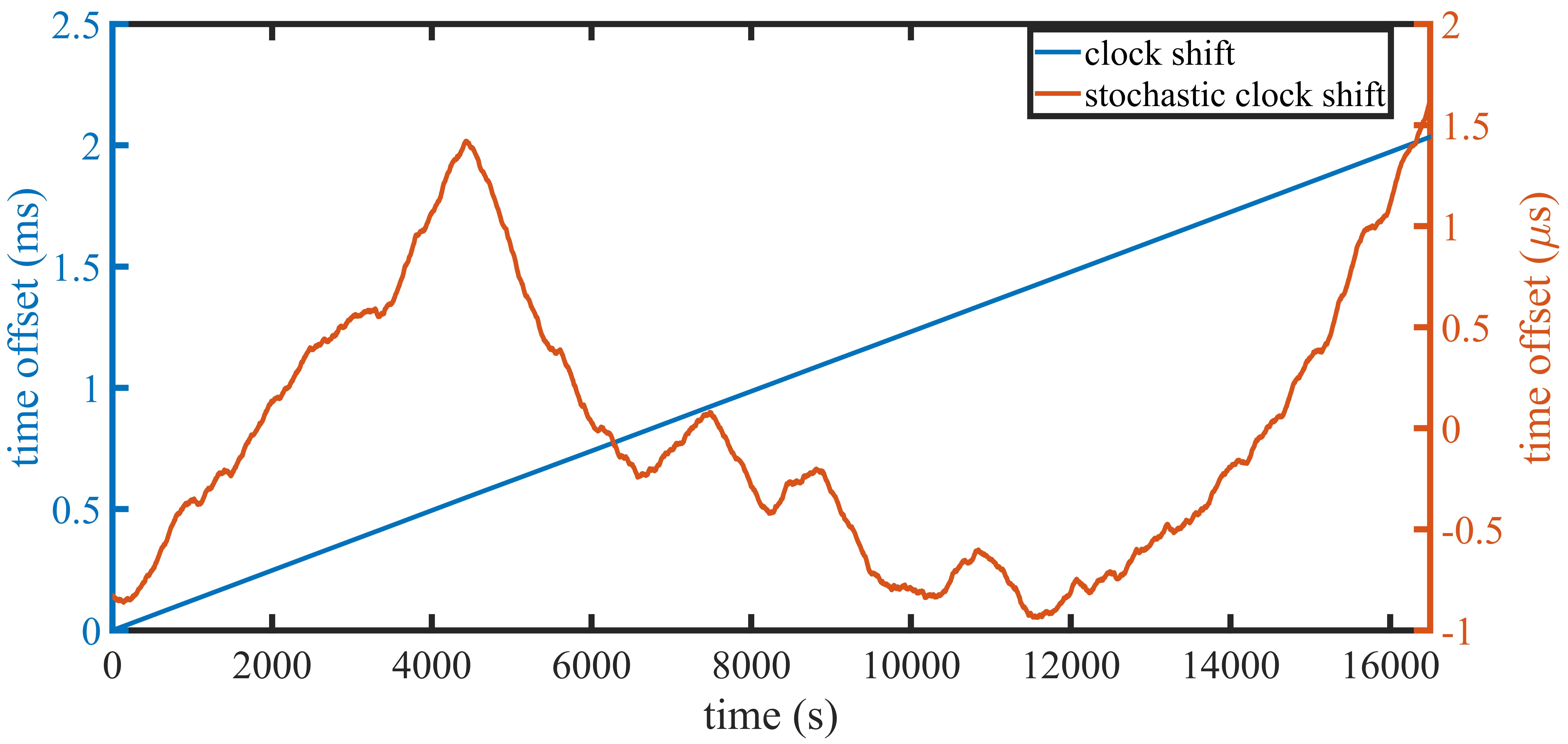}}
\caption{Clock shift between clock1 and clock2. The blue line shows the total time shift, the orange line shows the detrended time shift.}
\label{fig:2}
\end{figure}
\begin{figure}[ht]
\centering
{\includegraphics[width=12cm]{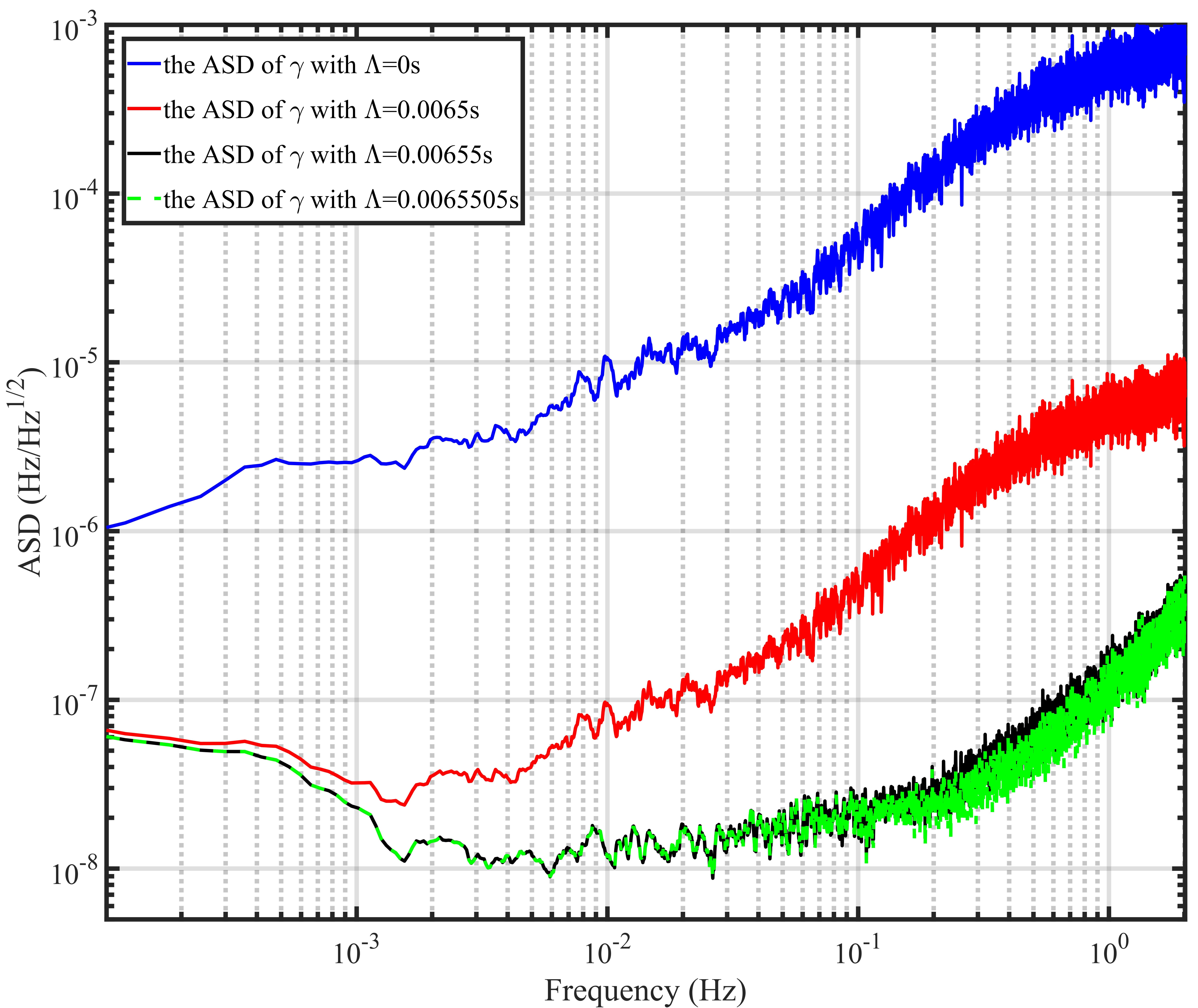}}
\caption{The results of clock synchronization of $\gamma$ using TDIR process.}
\label{fig:3}
\end{figure}

Fig. \ref{fig:2} shows that the clock offset grows to 2 ms over 16000 s according to Eq. (\ref{eq111}) (blue line), and the orange line shows the detrended
clock timing noise. Additionally, Fig. \ref{fig:3} shows the Amplitude spectral density (ASD) of Eq. (\ref{eq11}) at different translation times. With the improvement of clock synchronization accuracy, the residual noise in $\gamma$ becomes less which ultimately limited by the noise floor of the transmission link. In our experiment, the transmission link noise floor is dominated by high frequency signal sources. Finally, the initial offset can be determined to sub-microsecond accuracy with $\delta \tau _{2,0}\rm\approx\rm0.0065505$ $s$.

\begin{figure}[ht]
\centering
{\includegraphics[width=12cm]{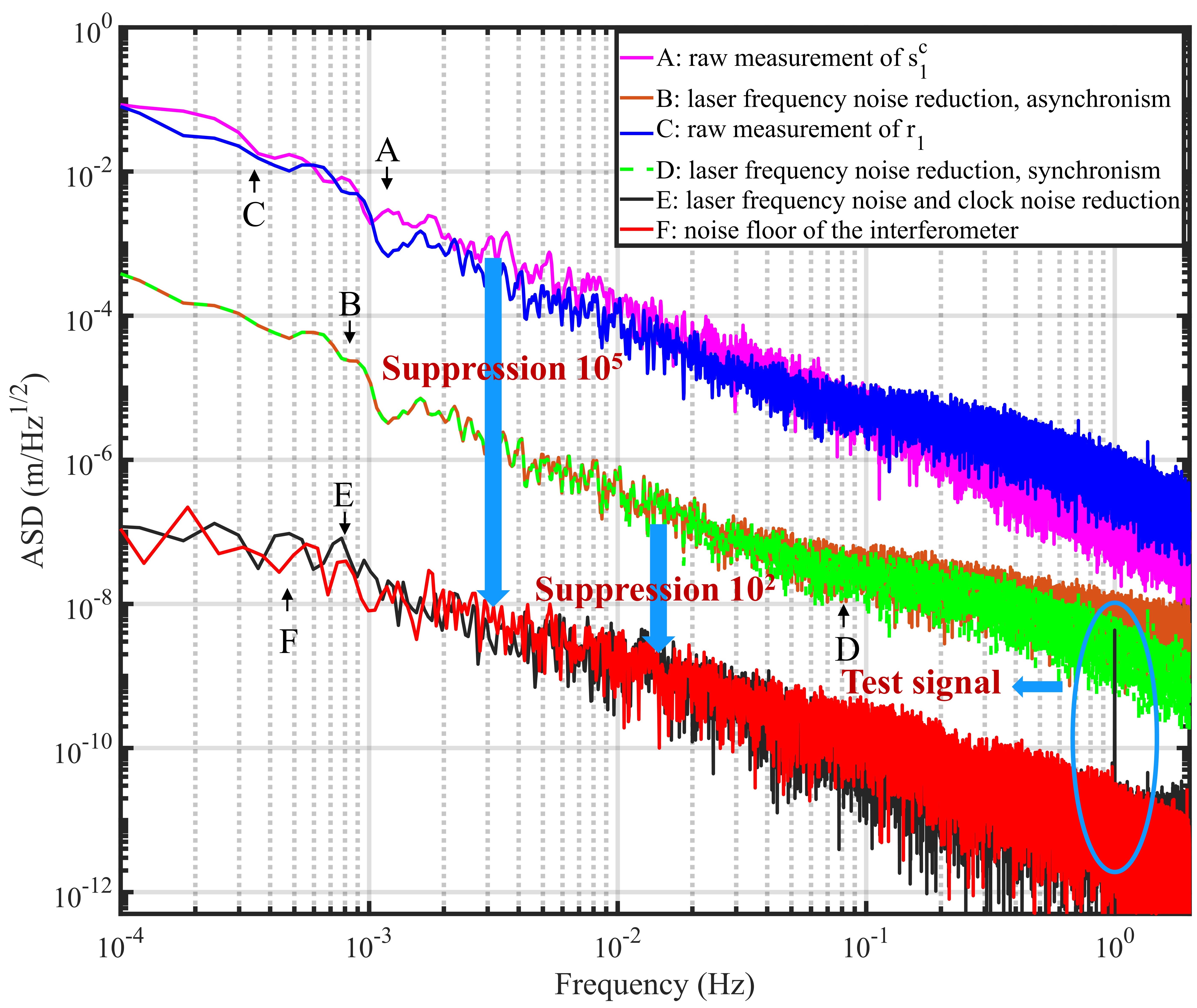}}
\caption{ASD of the results. A, raw measurement of $s_{1}^{c}$; B, linear combination of the two carrier data streams before clock synchronization; C, raw measurement of $r_1$; D, linear combination of the two carrier data streams after clock synchronization; E, processed data of $\eta$. }
\label{fig:4}
\end{figure}

In our experiment, the data collected is the form of the frequency unit, in order to get the information of the displacement, the data in frequency unit is integrand to get the information of the phase. Then, by times $\frac{\lambda}{2\pi}$, where $\lambda$ is the wavelength of the laser, we can convert phase into displacement.  Fig. \ref{fig:4} shows the frequency domain results. The line A shows the raw measurement data $s_{1}^{c}$ which is dominated by laser frequency noise with $1\times10^{-7}$ $\rm m/$ $\rm Hz^{1/2}$ at 1 $\rm Hz$ and $2\times10^{-3}$ $\rm m/$ $\rm Hz^{1/2}$ at 1 $\rm mHz$. The line B shows the result of eliminate laser frequency noise before clock synchronism with $s_{1}^{c}(t_1)-s_{2}^{c}(t_2)$ which is dominated by clock noise and residual laser frequency noise due to the clock asynchronism. The line D shows the result of eliminate laser frequency noise with clock synchronism as shown in Eq. (\ref{eq8}) with $3\times10^{-9}$ $\rm m/$ $\rm Hz^{1/2}$ at 1 $\rm Hz$ and $1\times10^{-5}$ $\rm m/$ $\rm Hz^{1/2}$ at 1 $\rm mHz$. Below 10 mHz, the line B and line D are almost same which is mainly limited by clock noise. Above 10 mHz, the noise of line B is higher than line D, which is due to the residual laser frequency noise caused by clock asynchronism. Since the clock asynchronism is mainly caused by the initial clock difference, and the difference between the two clocks is small, this clock asynchronism has less effect on the long time, mainly affecting the high frequency. The line C is related to the clock noise which is multiplied by the signal sources, shown in Eq. (\ref{eq3}). The lines C and D are trending in the same direction, indicating that they are affected by the same noise, i.e., the clock noise. Then, with Eq. (\ref{eq9}), We can suppress the clock noise below the interferometer noise floor, which is $1\times10^{-11}$ $\rm m/$ $\rm Hz^{1/2}$ at 1 $\rm Hz$ and $2\times10^{-8}$ $\rm m/$ $\rm Hz^{1/2}$ at 1 $\rm mHz$. The noise of the interferometer is mainly caused by the stability of the optical path and the fluctuation of temperature. Finally, the laser frequency noise and clock noise are suppressed by about 5 and 2 orders of magnitude and the weak displacement signal at 1 Hz can be seen as shown in line E. After 16000 s time accumulation, the ASD of this displacement signal reaches about $4\times10^{-9}$ $\rm m/$ $\rm Hz^{1/2}$ at 1 Hz. This is equivalent to a displacement signal with an average amplitude of 60 pm period 1s.
 \begin{figure}[!]
\centering
{\includegraphics[width=12cm]{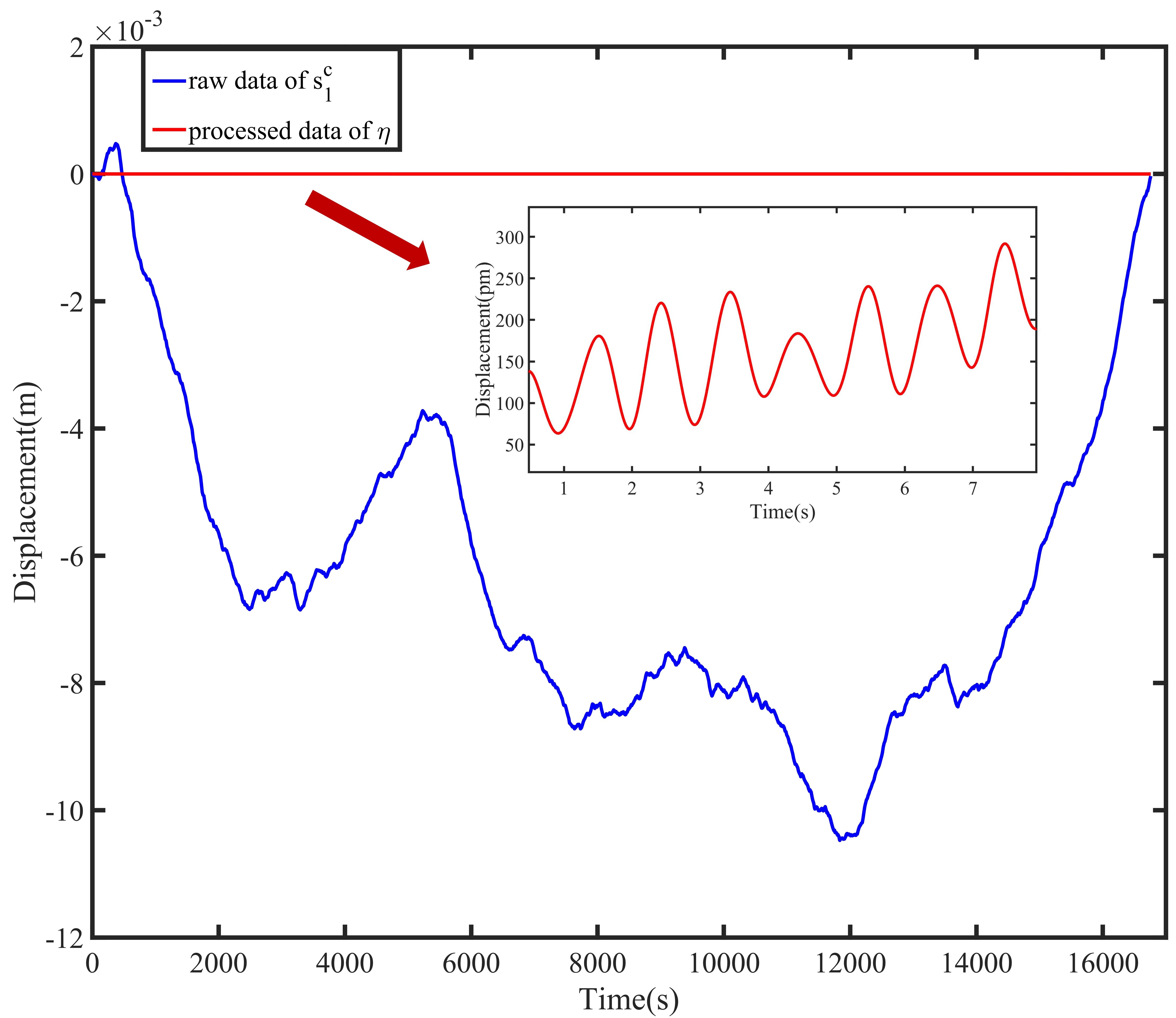}}
\caption{Time domain traces of the raw data and processed data. The blue line shows the raw displacement noise of $s_{1}^{c}$, the red line shows the displacement noise of $\eta$. A section of the trace is magnified to emphasize the displacement signal obtained in line red.}
\label{fig:5}
\end{figure}

Fig. \ref{fig:5} shows the time domain of raw data of $s_{1}^{c}$ and the processed data of $\eta$. The blue line shows the random laser frequency drift and clock drift, drowning out the signal. After data process, we can
see that the noise in the blue line was largely suppressed as shown in the red line. The inset in Fig. \ref{fig:5} enlarges a small section of the red line,  a displacement signal with average amplitude about 60 pm and period 1 s can be clearly seen.

\section{Conclusion}
In the future space-borne GW detection, the sensitive signal is extremely weak, and can be easily drowned out by noises. Laser frequency noise and clock noise are dominate noises which are introduced by unequal arm length and digital sampling process of the heterodyne interference signal. TDI post-processing techniques are used to eliminate them. In our experiment, although no delay is introduced, the idea of TDI constructing equal arms and the idea of clock sideband transfer comparison are used. The laser frequency noise is suppressed by 5 orders of magnitude, the clock noise is suppressed by 2 orders of magnitude, and finally, a 60 pm signal is restored, which is limited by the noise floor of the interferometer. The experimental results show that if the signal really exists, with the main noise being suppressed, the weak signal can be restored. At present, the system is mainly limited by interferometer noise. In the future, we will build an integrated optical bench to suppress optical noise. In addition, signal sources that multiply the clock frequency to GHz also introduce additional noise, and we will also focus on this electrical noise.

\section*{CRediT authorship contribution statement}
\textbf{Mingyang Xu}: Investigation, Methodology, Formal analysis, Writing – original draft, Writing – review $\&$ editing.
\textbf{Yujie Tan}: Conceptualization, Funding acquisition, Writing – review $\&$ editing.
\textbf{Yurong Liang}: Methodology, Funding acquisition, Writing – review $\&$ editing.
\textbf{Jiawen Zhi}: Formal analysis.
\textbf{Xiaoyang Guo}:Methodology.
\textbf{Dan Luo}: Formal analysis.
\textbf{Panpan Wang}: Investigation, Validation, Writing – review $\&$ editing.
\textbf{Hanzhong Wu}: Investigation, Methodology, Funding acquisition, Writing – review $\&$ editing.
\textbf{Cheng-Gang Shao}: Validation, Writing – review $\&$ editing, Funding acquisition, Project administration, Supervision.

\section*{Declaration of Competing Interest}
The authors declare no conflicts of interest.

\section*{Data availability}
 Data underlying the results presented in this paper are not publicly available at this time but may be obtained from the authors upon reasonable request.
 
\section*{Acknowledgments}
This work is supported by National Key Research and Development Program of China (2022YFC2204601, 2022YFC2203903); National Natural Science Foundation of China (11925503, 12275093 and 12175076); Natural Science Foundation of Hubei Province (2021CFB019), and State Key Laboratory of applied optics (SKLAO2022001A10).


  \bibliographystyle{elsarticle-num} 
  \bibliography{sample}





\end{document}